\documentclass{epl}

\usepackage{amsfonts}
\usepackage{theorem}
\usepackage{amssymb}
\usepackage{amsmath}
\usepackage{graphicx}

\newcommand{\rstr}{{\!\hbox{
$\vert\mkern-4.8mu\hbox{\rm\`{}}\mkern-3mu$}}}

\title{
Multi-distributed entanglement in finitely correlated
chains}\shorttitle{Distributed entanglement in chains}
\author{F.~Benatti\inst{1}\inst{2}, B.C.~Hiesmayr\inst{3} and H.~Narnhofer\inst{3}}
\institute{\inst{1} Dipartimento di Fisica Teorica, Universit\`a di Trieste,
Strada Costiera 11, \small 34014 Trieste, Italy\\
\inst{2}Istituto Nazionale di Fisica Nucleare, Sezione di Trieste, 34100
Trieste, Italy\\
\inst{3}Institut f\"ur Theoretische Physik, Boltzmanngasse 5, A-1090 Vienna,
Austria}


\begin{document}
 \maketitle

\begin{abstract}
The entanglement-sharing properties of an infinite spin-chain are studied when
the state of the chain is a pure, translation-invariant state with a
\textsf{matrix-product} structure~\cite{mat-prod}. We study the entanglement
properties of such states by means of their \textsf{finitely correlated
structure}~\cite{FNW}. These states are recursively constructed by means of an
auxiliary density matrix $\rho$ on a matrix algebra $\mathcal{B}$ and a
completely positive map
$\mathbb{E}:\mathcal{A}\otimes\mathcal{B}\to\mathcal{B}$, where $\mathcal{A}$
is the spin $2\times 2$ matrix algebra. General structural results for the
infinite chain are therefore obtained by explicit calculations in (finite)
matrix algebras. In particular, we study not only the entanglement shared by
nearest-neighbours, but also, differently from previous works \cite{Woo2}, the
entanglement shared between connected regions of the spin-chain. This range of
possible applications is illustrated and the maximal concurrence ${\cal
C}=\frac{1}{\sqrt{2}}$ \cite{Coff} for the entanglement of connected regions
can actually be reached.
\end{abstract}

\noindent The recent developments in quantum information and computation theory
are witnessing an increasing interest in the entanglement properties of
multi-qubit systems like quantum spin-chains, correlated electrons and
interacting bosons. These systems had so far been studied in condensed matter
physics especially in relations to their critical behaviour in
phase-transitions~\cite{Gold}. With its emergence as a precious resource for
quantum computation and communication tasks, entanglement is now investigated
in condensed matter physics, too; on one hand to detect, extract and manipulate
it, on the other hand to understand its role in solid state
phenomena~\cite{Ocon,Arne,Gunl,Zana,Gu,Osbo,Nar1,Nar2}.


The entanglement properties of the class of states of quantum spin
chains known as \textsf{Matrix-Product States}
(\textsf{MPS})~\cite{mat-prod}, have been proved powerful tools in
{\it density matrix renormalization groups techniques}~\cite{ZW-Vid}
and in universal quantum computation to show the equivalence of
teleportation-based and one-way quantum computation~\cite{Ver-Cir}.
They can be also achieved by sequential generation~\cite{SSVCW}.
Actually, these states were originally introduced in~\cite{FNW} as
\textsf{Finitely Correlated States} (\textsf{FCS}) and their
properties were there studied in great detail; in particular, it was
showed that \textsf{FCS} can be implemented as ground states of
Hamiltonians with short range interaction (see also~\cite{Ver-Cir}).
This motivates the study of entanglement in \textsf{FCS} both from
the point of view of the large variety of scaling laws provided and
from the point of view of the possible physical applications.


Two main different approaches have emerged: one dealing with the
scaling behaviour of entanglement-sharing  between various subsets
of localized spins~\cite{Fan,LRV,Oste,Subr}, the other one studying
how much entanglement can be localized on two distant spins by
measuring the others~\cite{Ver1,Ver2}.
In this Letter we choose the first approach and study the
entanglement of \textsf{FCS} by means of their recursive
structure~\cite{FNW}, that is by studying the specific completely
positive maps between finite dimensional algebras and finite
dimensional density matrices on which it is based. This provides a
rich phenomenology of total translation invariant states over
infinite spin chains whose entanglement properties are nevertheless
determined by, and can consequently be studied in a
finite-dimensional setting. Thus, we are able to compare nearest
neighbours with non-nearest neighbours entanglement or, more
generally, to analytically study the entanglement between different
subsets of lattice points. In particular, we shall give a general
necessary condition for one spin being entangled with a subset of
others and prove that in some cases this condition is also
sufficient. Finally, we show that entanglement sharing can achieve
its maximum in the sense of~\cite{Coff}.
\smallskip

\noindent
\textbf{Construction of translation invariant \textsf{FCS}:}\quad
We will denote by $\mathcal{A}_\mathbb{Z}$ an infinite spin-chain, the
spins at sites $i\in \mathbb{Z}$ being described by the algebra
$({\cal A})_i={\bf M}_2$ of $2\times 2$ complex matrices.
The infinite algebra $\mathcal{A}_\mathbb{Z}$ arises as a suitable
limit of the {\it local} tensor-product algebras
${\cal A}_{[-n,n]}:=\otimes_{j=-n}^n(\mathcal{A})_j$.
Any state $\omega$ over $\mathcal{A}_\mathbb{Z}$ is specified by
density matrices $\rho_{[1,n]}$ defining the action of $\omega$
as an expectation over local operators
$A_{[1,n]}\in\mathcal{A}_{[1,n]}$:
\begin{equation}
\label{fcs0}
\omega(A_{[1,n]})= {\rm Tr}_{[1,n]}\Bigl(\rho_{[1,n]}A_{[1,n]}\Bigr) \ .
\end{equation}
The $\rho_{[1,n]}$'s must satisfy the compatibility conditions
\begin{equation}
\label{comp}
\hbox{Tr}_{n+1}\Bigl(\rho_{[1,n+1]}A_{[1,n]}\otimes 1_{n+1}\Bigl)
=\hbox{Tr}\Bigl(\rho_{[1,n]}A_{[1,n]}\Bigr)\ ,
\end{equation}
whereas, translation-invariance requires
\begin{equation}
\label{trans-inv} \hbox{Tr}_{n+1}\Bigl(\rho_{[1,n+1]}1_1\otimes
A_{[2,n+1]}\Bigl) =\hbox{Tr}\Bigl(\rho_{[1,n]}A_{[1,n]}\Bigr)\ .
\end{equation}
The class of translation-invariant \textsf{FCS} over
$\mathcal{A}_{\mathbb{Z}}$ is defined by a triple
$(\mathcal{B},\rho,\mathbb{E})$ where $\mathcal{B}$ is a $b\times b$
matrix algebra $\mathcal{B}$,  $\rho\in\mathcal{B}$ a density matrix and
$\mathbb{E}:\mathcal{A}\otimes\mathcal{B}\mapsto\mathcal{B}$
a completely positive unital map, which in Kraus-Stinespring form reads
\begin{equation}
\label{Vop0}
\mathbb{E}(A\otimes B)=\sum_jV_j(A\otimes B)V_j^\dagger\ ,\quad
V_j:\mathbb{C}^2\otimes\mathbb{C}^b\mapsto\mathbb{C}^b\ ,
\end{equation}
with $A\in\mathcal{A}$ and $B\in\mathcal{B}$. \textbf{Unitality}
means that identities are preserved:
$\mathbb{E}(1_\mathcal{A}\otimes1_\mathcal{B})=1_\mathcal{B}$.

Let $\mathbb{E}^{(1)}(A):=\mathbb{E}(A\otimes1_\mathcal{B})$; this
defines a completely positive map from $\mathcal{A}$ into
$\mathcal{B}$.
Analogously, the recursive compositions
$\mathbb{E}^{(n)}:=\mathbb{E}\circ\Bigl({\rm id}_\mathcal{A}\otimes
\mathbb{E}^{(n-1)}\Bigr)$ are completely positive maps from
$\mathcal{A}_{[1,n]}$ into $\mathcal{B}$.
Setting
\begin{equation}
\label{st1}
\hbox{Tr}
\Bigl(\rho_{[1,n]}\,A_{[1,n]}\Bigr):={\rm Tr}_{\mathcal{B}}
\Bigl(\rho\,\mathbb{E}^{(n)}\Bigl(A_{[1,n]}\Bigr)\Bigr)\ ,
\end{equation}
the r.h.s. recursively defines local density matrices $\rho_{[1,n]}$
over $\mathcal{A}_{[1,n]}$ and a total state $\omega$ on
$\mathcal{A}_\mathbb{Z}$~\cite{FNW}. Further, ${\rm
Tr}_\mathcal{B}\Bigl(\rho\,\mathbb{E}\Bigl(1_{\mathcal{A}}\otimes
B\Bigr))\Bigr)={\rm Tr}_\mathcal{B}(\rho\,B)$, $\forall\,
B\in\mathcal{B}$ yields \textbf{translation-invariance}~\cite{FNW}.

Concretely, we choose
$\mathcal{B}={\bf M}_2$ and $\mathbb{E}$ in~(\ref{Vop0}) with just one Kraus
operator $V:\mathbb{C}^2\otimes\mathbb{C}^2\mapsto\mathbb{C}^2$.
This is such that $V\vert\phi_i\otimes\psi\rangle=v_i\vert\psi\rangle$,
$V^\dagger\vert\psi\rangle =\sum_{i=1}^2\vert\phi_i\rangle\otimes
v_i^\dagger\vert\psi\rangle$ and
\begin{equation}
\label{1kraus}
\mathbb{E}\Bigl(\vert\phi_i\rangle\langle\phi_j\vert\otimes B\Bigr)=
v_i\, B\, v_j^\dagger\ ,\quad B\in\mathcal{B}\ ,
\end{equation}
where $\vert\phi_{1,2}\rangle\in\mathbb{C}^2$ are orthonormal and
$v_{1,2}$ are $2\times 2$ matrices satisfying
\begin{eqnarray}\label{condition1}
v_1 v_1^\dagger+v_2 v_2^\dagger=1_\mathcal{B}\quad
(\hbox{\bf unitality})\ ,\quad
\sum_{j=1}^2v_j^\dagger\ \rho\ v_j=\rho\quad
(\hbox{\bf translation invariance})\;.
\end{eqnarray}
If there exists a unique $\rho$ fulfilling the previous condition,
the resulting translation--invariant \textsf{FCS} are pure states
over $\mathcal{A}_\mathbb{Z}$~\cite{FNW}, namely they cannot be
decomposed as mixtures of other states. These pure states can be
interpreted as ground states for appropriate constructed
Hamiltonians of finite range, but higher correlations~\cite{FNW}.

Consider~(\ref{st1}) with $n=2$ and $A_{[1,2]}=A_1\otimes A_2$, then
\begin{equation}
\label{expl}
\hbox{Tr}
\Bigl(\rho_{[1,2]}\,A_1\otimes A_2\Bigr):={\rm Tr}_{\mathcal{B}}
\Bigl(\rho\,\mathbb{E}\Bigl(A_1\otimes\mathbb{E}\Bigl(A_2\otimes1_\mathcal{B}
\Bigr)\Bigr)\ .
\end{equation}
Using the properties of the
trace-operation, the action of $\mathbb{E}$ becomes the
action of its dual map $\mathbb{F}$ onto the state $\rho\in\mathcal{B}$:
${\rm Tr}_\mathcal{B}(\rho\; \mathbb{E}(A\otimes B))
={\rm Tr}_{\mathcal{A}\otimes\mathcal{B}}(\mathbb{F}(\rho)\;A\otimes B)$.
This provides a state $\rho_{\mathcal{A}\otimes\mathcal{B}}:=
\mathbb{F}(\rho)=V^\dagger\,\rho\,V$ on $\mathcal{A}\otimes\mathcal{B}$:
\begin{equation}
\label{qbc10}
\rho_{\mathcal{A}\otimes\mathcal{B}}=
\sum_{s,t=1}^2\vert\phi_s\rangle\langle\phi_t\vert\otimes
v^\dagger_s\,\rho\, v_t
=\begin{pmatrix}
v_1^\dagger\rho v_1&v_1^\dagger\rho v_2\cr v_2^\dagger\rho
v_1&v_2^\dagger\rho v_2
\end{pmatrix}\ .
\end{equation}
The r.h.s. of~(\ref{expl}) reads
$\hbox{Tr}_{\mathcal{A}\otimes\mathcal{B}}
\Bigl(\rho_{\mathcal{A}\otimes\mathcal{B}}\,A_1\otimes\Bigl(\mathbb{E}
\Bigl(A_2\otimes 1_\mathcal{B}\Bigr)\Bigr)\Bigr)$, by
turning ${\rm id}_\mathcal{A}\otimes\mathbb{E}$
into its dual, nearest-neighbours states arise as
$\rho_{12}:=\rho_{[1,2]}
=\hbox{Tr}_\mathcal{B}\Bigl({\rm id}_\mathcal{A}\otimes\mathbb{F}
(\rho_{\mathcal{A}\otimes\mathcal{B}})\Bigr)$ and read
\begin{eqnarray}
\nonumber
\rho_{12}&=&\sum_{ijlm=1}^2
\vert\phi_i\rangle\langle\phi_j\vert\otimes
\begin{pmatrix}
R_{1ij1}&R_{1ij2}\cr
R_{2ij1}&R_{2ij2}\end{pmatrix}\\
\label{qbc11}
&=&
\begin{pmatrix}
R_{1111}&R_{1112}&R_{1121}&R_{1122}\cr
R_{2111}&R_{2112}&R_{2121}&R_{2122}\cr
R_{1211}&R_{1212}&R_{1221}&R_{1222}\cr
R_{2211}&R_{2212}&R_{2221}&R_{2222}\cr
\end{pmatrix}\ ,
\end{eqnarray}
where $R_{ijlm}={\rm Tr}(v_i^\dagger v_j^\dagger\; \rho\; v_l v_m)$,
while general local density matrices are given by
\begin{equation}
\label{qbc5}
\rho_{[1,n]}=\sum_{{\bf s},{\bf t}}\vert{\phi_{{\bf s}}}\rangle
\langle\phi_{{\bf t}}\vert\, {\rm Tr}(v_{{\bf
s}}^\dagger\,\rho\,v_{{\bf t}})\ ,
\end{equation}
where $\vert{\phi_{{\bf s}}}\rangle
=\vert\phi_{s_1}\otimes\phi_{s_2}\otimes\cdots\phi_{s_n}\rangle$,
$v_{\bf t}:=v_{t_1}\cdots v_{t_n}$.
\smallskip

We shall now relate the entanglement of
$\rho_{\mathcal{A}\otimes\mathcal{B}}$ to that of the spin at site $1$
with those at sites in $[p,n]$, $p>1$. This means
investigating the entanglement of the restricted state
$\omega\rstr(\mathcal{A})_1\otimes\mathcal{A}_{[p,n]}$.
By the previous construction, the restriction amounts to the
expectations
\begin{equation}
\omega\Bigl(A_1\otimes 1_{[2,p-1]}
\otimes A_p\otimes\cdots A_n\Bigr)
 \label{qbc7} ={\rm Tr}_{\mathcal{A}\otimes\mathcal{B}}
\Bigl(\rho_{\mathcal{A}\otimes\mathcal{B}}\
A_1\otimes\mathbb{G}(A_p\otimes\cdots A_n)\Bigr)\ ,
\end{equation}
where $1_{[2,p-1]}=(1)_2\otimes\cdots (1)_{p-1}$ and $\mathbb{G}$ is
a completely positive map from $\mathcal{A}_{[p,n]}$ into
$\mathcal{B}$. Notice that while $\rho_{12}$ is a nearest-neighbours
state, $\rho_{\mathcal{A}\otimes\mathcal{B}}$ encodes the
entanglement of the spin at site $1$ with any subset of spins
$\mathcal{A}_{[p,n]}$, $p>1$, after being embedded into
$\mathcal{B}=\mathcal{A}={\bf M}_2$ by $\mathbb{G}$. The advantage
of the abstract structure presented above emerges in that the total
state over the chain is determined by the triple
$(\mathcal{B},\mathbb{E},\rho)$, so that properties like the maximal
entanglement shared by nearest neighbours in a translation invariant
chain,  and, more generally, the entanglement between different
subsets of lattice points, can be studied by means of that triple
only.
\smallskip

\noindent
\textbf{Distribution of the entanglement along the chain:}\quad
As a measure of the entanglement of a state $\nu_{12}$ over the
tensor product algebra $\mathcal{N}_1\otimes\mathcal{N}_2$, we shall
use the entanglement of formation~\cite{Benn}
\begin{equation}
\label{entf} E_{\mathcal{N}_1\otimes\mathcal{N}_2}(\nu_{12})=
\inf_{\nu_{12}=\sum_j\lambda_j\nu^j_{12}} \Bigl\{ \sum_j\lambda_j\,
S\Bigl(\nu^j_{12}\rstr\mathcal{N}_1\Bigr) \Bigr\}\ ,
\end{equation}
where $\nu_{12}=\sum_j\lambda_j\nu^j_{12}$ is a convex decomposition
into pure states and $S\Bigl(\nu^j_{12}\rstr\mathcal{N}_1\Bigr)$ the
von Neumann entropy of their restrictions to the subalgebra
$\mathcal{N}_1$, namely of the state obtained by tracing over the
second factor, the density matrix representing the expectation
functional $\nu_{12}$. It turns out that
\begin{equation}
\label{monoton}
E_{(\mathcal{A})_1\otimes(\mathcal{A})_p}(\omega)\leq
E_{(\mathcal{A})_1\otimes\mathcal{A}_{[p,n]}}(\omega)\leq
E_{\mathcal{A}\otimes\mathcal{B}}(\rho_{\mathcal{A}\otimes\mathcal{B}})\ .
\end{equation}
The proof of this fact follows from a slight generalization of an argument
in~\cite{Nar3}. Let $\mathcal{N}_3$ be
another algebra, $\nu_{13}$ a state over
$\mathcal{N}_1\otimes\mathcal{N}_3$ and
$\Gamma:\mathcal{N}_2\mapsto\mathcal{N}_3$ a unital
($\Gamma(1_{\mathcal{N}_2})=1_{\mathcal{N}_3}$) completely positive map.
Then, $\nu_{12}=\nu_{13}\circ({\rm
id}_{\mathcal{N}_1}\otimes\Gamma)$ is a state over
$\mathcal{N}_1\otimes\mathcal{N}_2$, where $\circ$ denotes the
composition of the expectation functional $\nu_{13}$ with the
completely positive unital map ${\rm
id}_{\mathcal{N}_1}\otimes\Gamma$. As the
decompositions of $\nu_{12}$ induced by those of $\nu_{13}$ are not
all possible ones, it follows that
$E_{\mathcal{N}_1\otimes\mathcal{N}_2} (\nu_{12})$ is bounded from above by
\begin{equation}
\inf_{\nu_{13}=\sum_j\lambda_j\nu_{13}^j}\Bigl\{
\sum_j\lambda_j\, S\Bigl(\nu_{13}^j\circ({\rm id}_{\mathcal{N}_1}
\otimes\Gamma)\rstr\mathcal{N}_1\Bigr)\Bigr\}\ . \label{monoton0}
\end{equation}
From the unitality of $\Gamma$, the states $\nu_{13}^j\circ({\rm
id}_{\mathcal{N}_1} \otimes\Gamma)$ restricted to $\mathcal{N}_1$,
that is evaluating mean values of operators of the form
$N_1\otimes1_{\mathcal{N}_2}$, coincide with the restrictions of
$\nu_{13}^j$ themselves. Thus,
$E_{\mathcal{N}_1\otimes\mathcal{N}_2}(\nu_{12})\leq
E_{\mathcal{N}_1\otimes\mathcal{N}_3}(\nu_{13})$. The first
inequality in the hierarchy~(\ref{monoton}) follows by taking
$\mathcal{N}_1=(\mathcal{A})_1$, $\mathcal{N}_2=(\mathcal{A})_{p}$,
$\mathcal{N}_3=\mathcal{A}_{[p,n]}$ and as $\Gamma$ the natural
embedding of $(\mathcal{A})_p$ into $\mathcal{A}_{[p,n]}$. The
second one, by choosing $\mathcal{N}_1=(\mathcal{A})_1$,
$\mathcal{N}_2=\mathcal{A}_{[p,n]}$, $\mathcal{N}_3=\mathcal{B}$ and
$\Gamma=\mathbb{G}:\mathcal{A}_{[p,n]}\mapsto\mathcal{B}$; in fact,
$\omega\rstr(\mathcal{A})_1\otimes\mathcal{A}_{[p,n]}
=\rho_{\mathcal{A}\otimes\mathcal{B}}\circ{\rm
id}_\mathcal{A}\otimes\mathbb{G}$, see also Eq.~(\ref{qbc7}). Thus,
the general {\bf necessary condition} follows that, for
$(\mathcal{A})_1$ to be entangled with $\mathcal{A}_{[p,n]}$, $p>1$,
the state $\rho_{\mathcal{A}\otimes\mathcal{B}}$ must be entangled
over $\mathcal{A}\otimes\mathcal{B}$. Translation-invariance makes
this necessary condition independent of rigid shifts of the two
algebras; also, though the next examples go in the right direction,
unfortunately, there is so far no general argument for its
sufficiency namely that entanglement between $\mathcal{A}$ and
$\mathcal{B}$ should imply entanglement between $\mathcal{A}$ and
some $\mathcal{A}_{[p,n]}$.
\smallskip

\noindent \textbf{Examples:}\quad
Because $\mathcal{A}=\mathcal{B}={\bf M}_2$, we study the
{\it concurrence}~\cite{Woo1} of which the entanglement
of formation is a monotonically increasing function.
The concurrence of $\rho$
is given by $\mathcal{C}(\rho)=\max\lbrace
0,\lambda_1-\lambda_2-\lambda_3-\lambda_4\rbrace$, where $\lambda_j$
are the square roots of the eigenvalues in decreasing order of the
matrix $\rho\widetilde{\rho}$ where
$\widetilde{\rho}=(\sigma_y\otimes\sigma_y) \rho^*
(\sigma_y\otimes\sigma_y)$ and $\rho^*$ denotes complex conjugation
in the standard basis.
In terms of concurrence, inequality~(\ref{monoton}) reads
\begin{equation}
\label{monoton1}
\mathcal{C}\Bigl(\omega\rstr(\mathcal{A})_1\otimes(\mathcal{A})_p\Bigr)\leq
\mathcal{C}\Bigl(\omega\rstr(\mathcal{A})_1\otimes(\mathcal{A})_{[p,n]}\Bigr)
\leq\mathcal{C}\Bigl(\rho_{\mathcal{A}\otimes\mathcal{B}}\Bigr)\ .
\end{equation}

As a first finitely correlated structure, let us choose $v_1=
\begin{pmatrix}c&s\cr 0&0\end{pmatrix}$,
$v_2=\begin{pmatrix} 0&0\cr s&c\end{pmatrix}$, with $c=\cos\varphi$ and
$s=\sin\varphi$. Then the conditions (\ref{condition1}) are satisfied by
$\rho=1/2\begin{pmatrix}1&2cs\cr 2cs&1\end{pmatrix}$ and, using~(\ref{qbc5}),
the state of three adjacent qubits explicitly reads
\begin{eqnarray}
\label{sc1}
&&
\rho_{[1,3]}=\frac{1}{2}\sum_{i,j=1}^2\Omega_{ij}\otimes
\vert\phi_i\rangle\langle\phi_j\vert\otimes\Omega_{ij}\ ,\
\hbox{where}
\\
\label{sc1a}
&&
\Omega_{11}=\begin{pmatrix}c^2&2c^2s^2\cr
2c^2s^2& s^2
\end{pmatrix}\, ,\,
\Omega_{22}=\begin{pmatrix}s^2&2c^2s^2\cr
2c^2s^2& c^2
\end{pmatrix}\, ,\,
\Omega_{12}=\Omega_{21}^{\dagger}=
\begin{pmatrix}cs&2c^3s\cr
2cs^3& cs\end{pmatrix}\ ,
\end{eqnarray}
whereas the state $\rho_{\mathcal{A}\otimes\mathcal{B}}$  amounts to
\begin{equation}
\label{sc3} \rho_{\mathcal{A}\otimes\mathcal{B}}=\frac{1}{2}
\begin{pmatrix}
c^2& cs& 2c^2s^2 &2c^3s\cr
cs& s^2& 2cs^3& 2c^2s^2\cr
2c^2s^2& 2cs^3& s^2& cs\cr
2c^3s& 2s^2c^2& cs& c^2\cr
\end{pmatrix}\ .
\end{equation}
From~(\ref{sc1}), tracing over site $2$ and $3$, one gets the one-site
qubit-states $\rho_1=\frac{1}{2}\begin{pmatrix}1&4c^2s^2\cr 4c^2s^2&
1\end{pmatrix}$, while tracing over site $3$ we get the following two-sites
states,
\begin{equation}
\label{sc2}
\rho_{12}=\frac{1}{2}
\begin{pmatrix}c^2&2c^2s^2&2c^2s^2&4c^4s^2\cr
2c^2s^2& s^2&4c^2s^4&2c^2s^2\cr 2c^2s^2& 4c^2s^4& s^2 &2c^2s^2\cr 4c^4s^2&
2s^2c^2 &2c^2s^2 &c^2\cr
\end{pmatrix}
\end{equation}
and tracing over site $2$ gives
$\rho_{13}=\frac{1}{2}\Omega_{11}\otimes\Omega_{11}
+\frac{1}{2}\Omega_{22}\otimes\Omega_{22}$.

The latter state is clearly separable, there is no entanglement between site
$1$ and site $3$, i.e. no next nearest neighbour entanglement. While partially
transposing $\rho_{12}$, i.e. $4c^2s^4\leftrightarrow 4c^4s^2$ the positivity
is surely lost for $0\leq\varphi<\pi/4$, thus we have entanglement between
nearest neighbours. The same happens to the state of one site with the rest
$\rho_{\mathcal{A}\otimes\mathcal{B}}$, namely, if $s^2-4c^6<0$, that is for
$0\leq\varphi<\pi/4$, this state is surely not positive under partial
transposition.

A finer picture can be obtained by looking at the concurrences:
$\mathcal{C}_{\mathcal{A}\otimes\mathcal{B}}:=\mathcal{C}(\rho_{\mathcal{A}\otimes\mathcal{B}})
=\vert\sin{2\varphi}\cos(2\varphi)\vert$ vanishes at $\varphi=\pi/4, 5\pi/4$,
while $\mathcal{C}_{12}:=\mathcal{C}(\rho_{12})$, though explicitly computable,
is not as easily readable. Their behaviours as functions of
$\alpha=\sin(2\varphi)$ are reported in Figure 1(a) which shows firstly that,
in agreement with~(\ref{monoton1}),
$\mathcal{C}_{\mathcal{A}\otimes\mathcal{B}}\geq \mathcal{C}_{12}$ and secondly
that $\rho_{\mathcal{A}\otimes\mathcal{B}}$ is entangled whenever $\rho_{12}$
is entangled, namely for all $\sin(2\varphi)\neq0,\pm1$.

\begin{figure}
\center{
\includegraphics[width=150pt, keepaspectratio=true]{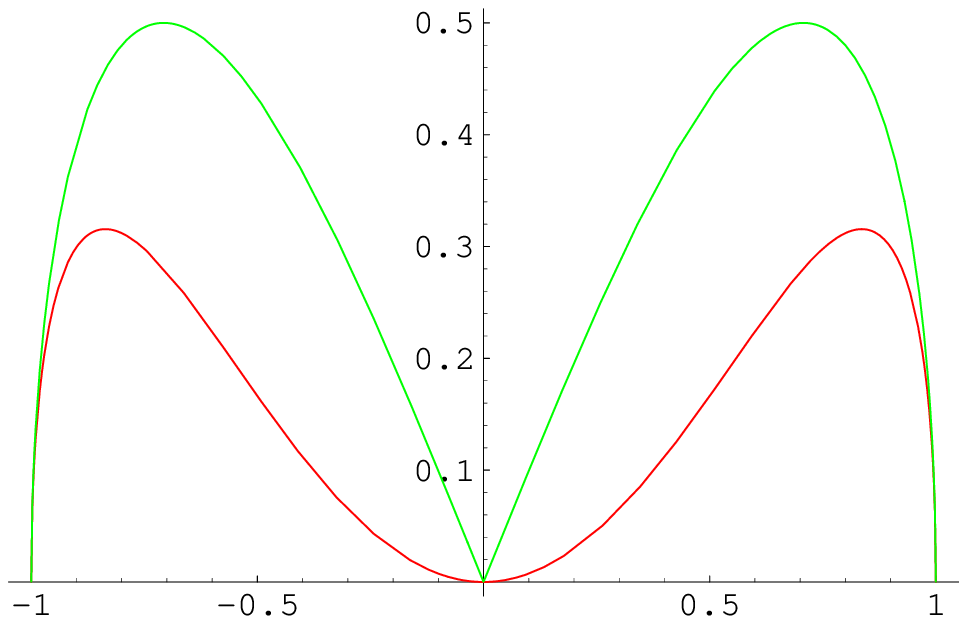}\hspace{3cm}
\includegraphics[width=150pt, keepaspectratio=true]{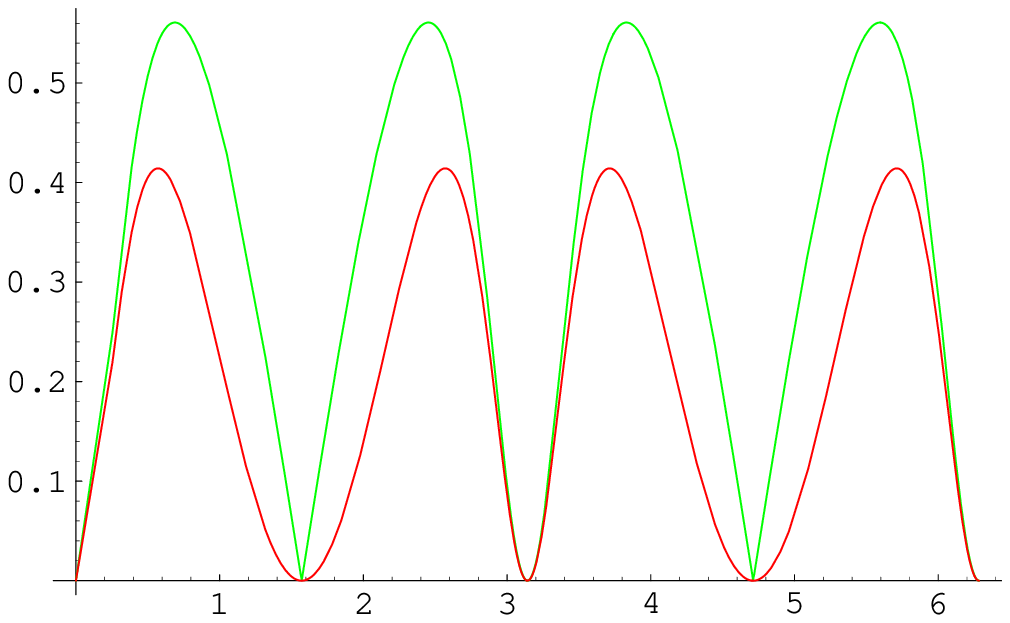}
\caption{(a) $\mathcal{C}_{\mathcal{A}\otimes\mathcal{B}}$, green;
$\mathcal{C}_{12}$, red; abscissa $\alpha=\sin(2\varphi)$. \hspace{0.5cm} (b)
$\mathcal{C}_{\mathcal{A}\otimes\mathcal{B}}$, green; $\mathcal{C}_{12}$, red;
abscissa $\varphi$.} \label{fig1}}
\end{figure}

Another finitely correlated state is given by choosing
\begin{equation}
\label{sd1}
v_1=\begin{pmatrix}
a c&a s\cr
-s& c
\end{pmatrix}\ ,\
v_2=\sqrt{1-a^2}\begin{pmatrix}
0&1\cr
0&0
\end{pmatrix}\ ,\ 0\leq a\leq 1\ ,
\end{equation}
with $s=\sin\varphi$ and $c=\cos\varphi$. Conditions (\ref{condition1}) give
\begin{equation}
\label{sd1a} \rho=\begin{pmatrix} x&y\cr y&1-x\end{pmatrix}\ ,\
\left\{\begin{matrix} x=\frac{s^2}{(1-a)^2 c^2\ +\ 2 s^2}\cr
y=\frac{(a-1) s c}{(1-a)^2 c^2\ +\ 2 s^2}
\end{matrix}\right.\ .
\end{equation}
With this choice $(v_2)^2=0$ and all terms in~(\ref{qbc11}) with two
adjacent $v_2$ or $v_2^\dagger$ vanish; therefore, independently of
$a,s$ in~(\ref{sd1a}),
nearest-neighbours states are of the form
\begin{equation}
\label{sd2} \rho_{12}=\begin{pmatrix}1-2 \gamma &\alpha&\alpha&0\cr
\alpha&\gamma&\beta&0\cr \alpha&\beta&\gamma&0\cr 0&0&0&0\cr
\end{pmatrix}\ ,
\end{equation}
and have concurrence $\mathcal{C}_{12}=2\vert \beta\vert$. The entanglement of
formation is maximal when
\begin{equation}
\label{sd2a}
\mathcal{C}_{12}=2{\rm Tr}(v_2^\dagger v_1^\dagger\rho v_2v_1)=
\frac{2 (1-a^2)s^2c^2}{(1-a)^2 c^2\ +\ 2 s^2}\ ,
\end{equation}
is maximal, that is when $a=\sqrt{2}-1$, $\cos(2 \phi_1)=\sqrt{2}-1$
and $\beta=\frac{1}{2}(\sqrt{2}-1)$, whence
$\mathcal{C}_{12}=\sqrt{2}-1=0.41$. On the other hand, the
concurrence of $\rho_{\mathcal{A}\otimes\mathcal{B}}$ amounts to
\begin{equation}
\label{sd2b} \mathcal{C}_{\mathcal{A}\otimes\mathcal{B}}=\frac{2\sqrt{1-a^2}
s^2 \vert c\vert} {(1-a)^2 c^2\ +\ 2 s^2}\ .
\end{equation}
In Figure 1(b), $a$ is set $\sqrt{2}-1$ and~(\ref{sd2a}--\ref{sd2b})
are plotted against $\varphi\in[0,2\pi]$: again agreement is shown
with the general monotonicity of concurrence expressed
by~(\ref{monoton1}). Namely, nearest-neighbours are entangled if the
state of one site with the rest
$\rho_{\mathcal{A}\otimes\mathcal{B}}$ is entangled, but we do not
know if entanglement of one site with rest always implies nearest
neighbours entanglement. We believe this to be peculiar of our
choice of $\mathcal{B}$: since it is two dimensional both
$\rho_{12}$ and $\rho_{\mathcal{A}\otimes\mathcal{B}}$ have rank $2$
and separable density matrices of rank $2$ have $0$ measure.

When three qubits share equal entanglement, then the maximum entanglement of
one qubit with each of the two other qubits is the sum of the squares of the
two concurrences which is less than or equal to one~\cite{Coff}. Thus nearest
neighbour concurrence cannot exceed $1/\sqrt{2}$, but it is an open problem
whether this upper bound can be achieved by an entangled chain. In~\cite{Woo2}
an entangled chain is considered in a translational-invariant state with
nearest-neighbours states as in~(\ref{sd2}) but with $\alpha=0$ and an upper
bound for shared entanglement corresponding to a greater concurrence
$\mathcal{C}_{12}=0.434467$.

The translation invariant state over the system
$\mathcal{A}_{[-\infty,-1]}\otimes\mathcal{A}_0\otimes\mathcal{A}_{[1,\infty]}$
can be defined with $\bar {\mathbb G}\otimes 1\otimes \mathbb{G}$ mapping the
system into $\mathcal{B}\otimes\mathcal{A}\otimes\mathcal{B}$. Now we have a
effectively three qubit state sharing equal amount of entanglement and due to
the considerations in \cite{Coff} the upper limit for the concurrence is given
by $\frac{1}{\sqrt{2}}$. In the following example, we show how the maximum can
indeed be reached by a finitely correlated structure. We set $v_1$ as
in~(\ref{sd1}), $v_2=\sqrt{1-a^2}\begin{pmatrix} 1&0\cr0&0\end{pmatrix}$ and
\begin{eqnarray}
\label{sdd1} && \rho=\frac{1}{1+a^2}\begin{pmatrix} 1&0\cr0&a^2\end{pmatrix}\
,\; \hbox{then}\quad
\rho_{\mathcal{A}\otimes\mathcal{B}}=\frac{\sqrt{1-a^2}}{1+a^2}\begin{pmatrix}
\frac{a^2}{\sqrt{1-a^2}}&0&ac&0\cr 0&\frac{a^2}{\sqrt{1-a^2}}&as&0\cr
ac&as&\sqrt{1-a^2}&0\cr 0&0&0&0
\end{pmatrix}\;,
\end{eqnarray}
whence $\displaystyle
C(\rho_{\mathcal{A}\otimes\mathcal{B}})=\frac{2a\sqrt{1-a^2}}{1+a^2}
\sin\varphi$
which attains its maximum $1/\sqrt{2}$ at $\varphi=\pi/2$ and $a=1/\sqrt{3}$.
This state differs from the one in Eq.~(\ref{sd2}). Therefore optimizing the
entanglement of $\mathcal{A}_0\otimes\mathcal{A}_{[1,\infty]}$ reduces the
entanglement of $\mathcal{A}_0\otimes\mathcal{A}_{1}$.
\smallskip

\noindent \textbf{Conclusions:}\quad We studied the entanglement
properties of \textsf{FCS} over infinite quantum spin chains by
means of their recursive structure and illustrated some possible
behaviors by examples with $2\times 2$ matrices. Investigation of
higher dimensional contexts as the AKLT model~\cite{AKLT} and
comparison with different approaches~\cite{Fan} is in progress.
\smallskip

\noindent  \textbf{Acknowledgment}:\quad
B.C. Hiesmayr acknowledges EURIDICE HPRN-CT-2002-00311.

\end{document}